\documentstyle[aps,eqsecnum,preprint,tighten]{revtex}

\def\beq{\begin{equation}}
\def\eeq{\end{equation}}
\def\bea{\begin{eqnarray}}
\def\eea{\end{eqnarray}}
\def\Otogamma{\Omega^- \rightarrow \Xi^- \gamma}
\def\to{\rightarrow}
\begin{document}
\draft
\preprint{\vbox{Submitted to {\it Physics Letters  \bf{B}}\hfill
IFT-P.061/95, IFUSP/P-1191\\}}
\title{ Can $\Omega^- \rightarrow \Xi^- \gamma$ be sensitive to new
physics beyond
the Standard Model ? }
\author{R. Rosenfeld}
\address{Instituto de F\'{\i}sica Te\'orica, Universidade Estadual Paulista \\
Rua Pamplona, 145; 01405-900 -- S\~ao Paulo,SP, Brazil }
\author{L.A. Barreiro, C.O. Escobar and M. Nielsen}
\address{Instituto de F\'{\i}sica, Universidade de S\~ao Paulo \\
Caixa Postal 66318 - 5389-970 S\~ao Paulo, SP, Brazil}
\maketitle
\begin{abstract}
We investigate the impact of new physics beyond the Standard Model to the
$s \rightarrow d \gamma$ process, which is responsible for the short-distance
contribution to the radiative decay $\Otogamma$. We study three representative
extensions of the Standard Model, namely a one-family technicolor model, a
two Higgs doublet model and a model containing scalar leptoquarks.
When constraints arising from the observed $b \to s \gamma$ transition and
the upper limit on  $D^0 - \bar{D}^0$ mixing are
taken into account, we find no significant contributions of new physics to
the $s \to d \gamma$ process.
\end{abstract}
\pacs{PACS numbers: }
\newpage

\section{Introduction}
It has been recently shown \cite{omega} that the short distance contribution to
the radiative decay $\Otogamma$, given by the penguin transition
$s \to d \gamma$ is comparable to the long-distance contributions
 \cite{Safadi-Singer}. This process is the second generation
analogue of the $b \rightarrow s \gamma$ transition recently
seen at CLEO which has
generated a large amount of interest \cite{bsg}. Being a one-loop diagram ,
it may be sensitive to
physics beyond the Standard Model not yet accessible by direct searches in
accelerators. The results from CLEO already put rather severe limits on various
models
for new physics \cite{topten}.
Another example of the impact of new physics on one-loop processes is the
case of rare top quark decays, which can have their branching ratios
enhanced by $3-4$ orders of magnitude \cite{raretop}.

At this moment, there is no observation of the transition
$s \rightarrow d \gamma$.
Experimentally, there is an upper limit determined recently \cite{ivone}:
\beq
BR^{Exp.}( \Omega^- \rightarrow \Xi^- \gamma) < 4.6  \times 10^{-4} .
\label{exp}
\eeq
The purpose of this work is to investigate whether this upper limit can
already provide useful constraints on physics beyond the Standard Model.

In ref. \cite{omega} we have estimated the short-distance
contribution to
the decay $\Otogamma$  using QCD sum rules for a more reliable
calculation of the non-perturbative contribution to the matrix element.
We have also re-estimated the Wilson coefficient
in the Standard Model that is responsible for the transition.
In this letter we examine the effects of new physics on the $s \to d \gamma$
transition. For definiteness, we focus on three representative examples
of new physics: a one-generation technicolor model, a two Higgs doublet model
and
 a model containing scalar leptoquarks. We compute the impact of these models,
taking into
consideration
the constraints on these models coming from the $b \to s \gamma$ transition
(for the first two models) and the upper limit on $D^0 - \bar{D}^0$
mixing (for the latter model).

The effective hamiltonian (with the heavy quarks $c,b$ and $t$ as well as
electroweak gauge bosons integrated out ) that describes $| \Delta S = 1 |$
transitions is given by :
\begin{equation}
{\cal H }_{\mbox{eff.}}^{| \Delta S = 1 | } = \frac{-4 G_F}{\sqrt{2}} \;
\lambda_u \; \sum_{k=1}^8 \; c_k(\mu) {\cal O}_k(\mu) \; ,
\label{H}
\end{equation}
where $G_F$ is the Fermi constant, and we use the notation
 $\lambda_i$ to
denote the following product of elements of the Cabibbo-Kobayashi-Maskawa (CKM)
matrix, $\lambda_i  =
V_{si} V^{\ast}_{id} $ .
The Wilson coefficients $c_k$ can be computed
perturbatively and $\{ {\cal O}\}_k$ is a complete set of operators written as
:
\begin{eqnarray}
{\cal O}_1 &=& (\bar{u}_{\alpha} \gamma_{\mu} P_L s_{\beta} )
              (\bar{d}_{\beta} \gamma^{\mu} P_L u_{\alpha} )  \\  \nonumber
{\cal O}_2 &=& (\bar{u}_{\alpha} \gamma_{\mu} P_L s_{\alpha} )
              (\bar{d}_{\beta} \gamma^{\mu} P_L u_{\beta} )  \\  \nonumber
{\cal O}_3 &=& (\bar{d}_{\alpha} \gamma_{\mu} P_L s_{\alpha} )
\sum_{q=u,d,s} (\bar{q}_{\beta} \gamma^{\mu} P_L q_{\beta} )  \\  \nonumber
{\cal O}_4 &=& (\bar{d}_{\alpha} \gamma_{\mu} P_L s_{\beta} )
\sum_{q=u,d,s} (\bar{q}_{\beta} \gamma^{\mu} P_L q_{\alpha} )  \\
\label{ope}
{\cal O}_5 &=& (\bar{d}_{\alpha} \gamma_{\mu} P_L s_{\alpha} )
\sum_{q=u,d,s} (\bar{q}_{\beta} \gamma^{\mu} P_R q_{\beta} )  \\  \nonumber
{\cal O}_6 &=& (\bar{d}_{\alpha} \gamma_{\mu} P_L s_{\beta} )
\sum_{q=u,d,s} (\bar{q}_{\beta} \gamma^{\mu} P_R q_{\alpha} )  \\  \nonumber
{\cal O}_7 &=& \frac{e}{16 \pi^2} m_s (\bar{d}_{\alpha} \sigma_{\mu \nu} P_R
              s_{\alpha} ) F^{\mu \nu}  \\  \nonumber
{\cal O}_8 &=& \frac{g_s}{16 \pi^2} m_s (\bar{d}_{\alpha} \sigma_{\mu \nu}
 T_{\alpha \beta}^a  P_R
              s_{\beta} ) G^{a \mu \nu}  \nonumber
\end{eqnarray}
where $ P_{\tiny{\begin{array}{l}
L\\R\end{array}}}=\frac{1}{2}(1\mp\gamma_5) $ , $\sigma_{\mu \nu} =
 i [ \gamma_{\mu}, \gamma_{\nu} ] /2 $; $F^{\mu \nu}$ and $G^{a \mu \nu}$ are
the electromagnetic and color field tensors respectively.

The Wilson coefficients are first computed at the scale $\mu = m_w$ in zeroth
order in QCD. QCD corrections are included by evolving these coefficients down
to $\mu = m_s$ using the renormalization group equations. In our case,
this evolution
is accomplished in three steps ($\mu = m_b,m_c$ and $m_s$) .

Our result, obtained by using the leading-logarithm anomalous dimension
matrix given
by Buras et al. \cite{Buras} is \cite{omega}  :
\begin{equation}
c_7(m_s) = -0.50 c_2(m_w) + 0.27 c_7(m_w) + 0.13 c_8(m_w)
\end{equation}
with the initial values for the Wilson coefficients at the $m_w$ scale in the
Standard Model given by :
\begin{eqnarray}
c_7^{SM}(m_w) &=& (2.1 - 7.7) \times 10^{-4} \\
c_8^{SM}(m_w) &=& (1.4 - 4.2) \times 10^{-4}  \\
c_2(m_w) &=& 1
\end{eqnarray}
where the uncertainties are due to the poorly determined CKM matrix
elements $V_{td}$ and $V_{ts}$ \cite{PDG}.

The small values of $c_7$ and $c_8$ reflect the large suppression due to
CKM matrix elements for the top-quark intermediate state, whereas the
contribution
from the charm-quark intermediate state is suppressed due to its small
mass.

New physics at energy scales larger that $m_w$ will manifest itself in the
initial
values of the Wilson coefficients $c_7$ and $c_8$ at the scale $m_w$.
We can write these coefficients in general as :
\begin{eqnarray}
c_7(m_w) &=& c_7^{SM}(m_w) + c_7^{NP}  \\
c_8(m_w) &=& c_8^{SM}(m_w) + c_8^{NP} ,
\end{eqnarray}
where $c_{7,8}^{NP}(m_w)$ represents the contributions from physics beyond the
Standard Model at the scale $m_w$.

One can see from equations ($1.5 - 1.10$) that enhancements of
roughly $4$ orders of magnitude with
respect to the Standard Model values are necessary
in order to make $c_{7}(m_s)$ sensitive to the initial values $c_{7,8}(m_w)$.

In the next sections we briefly describe the models chosen as representative
of New Physics and compute their contribution to $ c_7^{NP}$ and
$ c_8^{NP}$.

\section{One-Family Technicolor}

Although there isn't a realistic technicolor model for electroweak
symmetry breaking that is in agreement with precision measurements, the
general idea is still attractive and the one-family technicolor model, where
one
introduces a whole family of techni-fermions, has become the prototype
for these models \cite{techni}. The one-family technicolor model possesses
a global
$SU_L(8) \times SU_R(8)$ symmetry that is broken down to $SU_V(8)$ due to
techni-fermions condensation. In this process, $63$ pseudo-Nambu-Goldstone
bosons are generated, $3$ of which become the longitudinal components of
the electroweak gauge bosons. We will concentrate on the contribution from
the charged, colored pseudo-Nambu-Goldstone bosons ($P_8^{\pm}$),
which dominates over
the colorless ones, which is given by \cite{pgb} :

\begin{equation}
c_7^{TC}(y_q) = \frac{-8}{3 \sqrt{2} G_F F_{\pi}^2}  \; \left\{
\frac{ -\frac{11}{36} + \frac{53}{72} y_q - \frac{25}{72} y_q^2}{(1-y_q)^3}  +
\frac{ -\frac{1}{4} y_q + \frac{2}{3} y_q^2 - \frac{1}{3} y_q^3}{(1-y_q)^4}
\log[y_q]
\right\}
\end{equation}
and
\begin{equation}
c_8^{TC}(y_q) =  \frac{-1}{3 \sqrt{2} G_F F_{\pi}^2}  \; \left\{
\frac{ -\frac{1}{6} + \frac{107}{24} y_q - \frac{145}{24} y_q^2}{(1-y_q)^3}  +
\frac{ \frac{9}{4} y_q - \frac{5}{2} y_q^2 - 4 y_q^3}{(1-y_q)^4} \log[y_q]
\right\}
\end{equation}
where $y_q = (m_{P_8^{\pm}}/m_q)^2 $ and $F_{\pi} = 123$ GeV in the one-family
technicolor
model.
In order to obtain the final result we have to sum over the contributions from
the
charm and top-quarks intermediate states multiplied by the appropriate CKM
matrix elements :
\begin{eqnarray}
c_7^{TC} &=& \frac{\lambda_t}{\lambda_u} c_7^{TC}(y_t) +
\frac{\lambda_c}{\lambda_u}
c_7^{TC}(y_c)  \\
c_8^{TC} &=& \frac{\lambda_t}{\lambda_u} c_8^{TC}(y_t) +
\frac{\lambda_c}{\lambda_u}
c_8^{TC}(y_c)
\end{eqnarray}

The final result depends only on $m_{P_8^{\pm}}$ and in table I we show the new
coefficients for $m_{P_8^{\pm}} = 80 \;,\; 400$ GeV.
The value $m_{P_8^{\pm}} = 400$ GeV is the lower
 limit
on these models derived from the process $b \to s \gamma$ \cite{pgb}.
Ignoring such constraints, the minimum value of $m_{P_8^{\pm}}$ we could
introduce in the model without changing
the
renormalization group equations is $m_{P_8^{\pm}} = 80$ GeV.
In order to find the maximum allowed enhancements,
we have taken the largest possible value of the CKM matrix elements.

We conclude that in this model, the resulting enhancement is not large
enough to make $c_{7,8}(m_w)$ comparable to $c_2(m_w)$. Taking into account
the constraints from $b \to s \gamma$, only enhancements of a factor $10$
at most are allowed.

\section{Two Higgs Doublet Models}

Many extensions of the Standard Model predict the existence of two Higgs
doublets. The Minimal Supersymmetric Standard Model is a popular example.
Tree-level flavor changing neutral currents can be avoided in two classes of
models, commonly referred as I and II. In model I, one doublet ($\phi_2$)
provides masses to all fermion while the other doublet ($\phi_1$) decouples
from the fermion sector. In model II, $\phi_2$ gives masses to the up-type
quarks while $\phi_1$ gives masses to charged leptons and down-type quarks.
These models have charged Higgs scalars in their spectrum which contribute
to the $s \to d \gamma$ processes. Their contribution depends on the charged
Higgs boson mass ($M_{H^{\pm}}$) and on the ratio of the vacuum expectation
values of the two scalar fields $\phi_1$ and $\phi_2$, $\tan \beta = v_2/v_1$
and it is given by \cite{2hdm} :
\begin{eqnarray}
c_7^{2HDM}(y_q) &=& \cot^2 \beta \{ B(y_q) - A(y_q)/6 \}  \\
c_8^{2HDM}(y_q) &=& \cot^2 \beta \{ E(y_q) - D(y_q)/6 \}
\end{eqnarray}
where $y_q = m_q^2/M_{H^{\pm}}^2$ and
\begin{eqnarray}
A(x) &=& x \left\{ \frac{ \frac{2}{3} x^2 - \frac{5}{12} x - \frac{7}{12}
}{(x-1)^3}
- \frac{ \frac{3}{2} x^2 - x}{(x-1)^4} \log[x]   \right\}  \\
B(x) &=&  \frac{x}{2} \left\{ \frac{ \frac{5}{6} x - \frac{1}{2} }{(x-1)^2}
- \frac{ x - \frac{2}{3} }{(x-1)^3} \log[x]   \right\}  \\
D(x) &=&  \frac{x}{2} \left\{ \frac{ \frac{1}{2} x^2 - \frac{5}{2} x - 1
}{(x-1)^3}
+ \frac{ 3 x}{(x-1)^4} \log[x]   \right\}  \\
E(x) &=&  \frac{x}{2} \left\{ \frac{ \frac{1}{2} x - \frac{3}{2} }{(x-1)^2}
+ \frac{\log[x] }{(x-1)^3}  \right\}
\end{eqnarray}

{}From these expressions it is clear that the maximum enhancement will arise
from the small $\tan \beta$-- small $M_{H^{\pm}}$ region in the parameter
space.

As before, one should sum over the charm and top-quarks contributions :
\begin{eqnarray}
c_7^{2HDM} &=& \frac{\lambda_t}{\lambda_u} c_7^{2HDM}(y_t) + \frac{\lambda_c}
{\lambda_u}
c_7^{2HDM}(y_c)  \\
c_8^{2HDM} &=& \frac{\lambda_t}{\lambda_u} c_8^{2HDM}(y_t) + \frac{\lambda_c}
{\lambda_u}
c_8^{2HDM}(y_c)
\end{eqnarray}

In tables II and III we show upper values of $c_7^{2HDM}$ and $c_8^{2HDM}$,
taking
into account the constraints arising from $b \to s \gamma$ \cite{hewett}.

Therefore, in both models the enhancement are small given the constraints
from $b \to s \gamma$.

\section{Leptoquarks}

Leptoquarks are particles (scalar or vector) that carry both baryon and
lepton numbers and hence couple to a lepton and a quark. Their interactions
are described by renormalizable, baryon and lepton number conserving,
$SU(3)_c \times SU(2)_L \times U(1)_Y$ invariant lagrangians \cite{lepto1}.
Leptoquarks may arise in some extensions of the Standard Model, like
Grand Unified theories, Extended Technicolor models and models with quark
and lepton substructure \cite{lepto2}.

In this work we focus on the contributions of a scalar leptoquark to the
process $s \to d \gamma$, which can be written as \cite{lepto2} :
\begin{equation}
c_7^{LQ}(m_w) = - \frac{\lambda^2}{24 \sqrt{2} G_F m_{LQ}^2 \lambda_u} \;\;
		\left( \frac{Q_{LQ}}{2} + Q_L \right) ,
\end{equation}
where $\lambda$ represents a generic leptoquark coupling constant, $m_{LQ}$ and
$Q_{LQ}$ are
the leptoquark mass and electric charge respectively and $Q_L$ is the
electric charge of the
lepton in the loop.

The leptoquarks masses and couplings are restricted by a variety of processes
and here we take the coupling with the least stringent bound, $\lambda_{R S_0}$
($Q_{S_0} = -1/3$) in the notation of Davidson {\it et al.} \cite{lepto2} :
\begin{equation}
\lambda_{R S_0}^2 < 6 \times 10^{-3} \left( \frac{m_{LQ}}{100 GeV} \right)^2
\end{equation}
which arises from $D^0 - \bar{D}^0$ mixing.
In this case we arrive at :
\begin{equation}
|c_7^{LQ}(m_w)| < 6 \times 10^{-3}
\end{equation}
Therefore, even in the most optimistic case, the contributions from leptoquarks
are not sufficient to make the $s \to d \gamma$ process sensitive to their
existence.

\section{Conclusion}

We have studied the influence of three representative models of physics
beyond the Standard Model to the process $s \to d \gamma$, which is
responsible for the short distance contribution to $\Otogamma $ . Given the
constraints on these models arising from $b \to s \gamma$ and
$D^0 - \bar{D}^0$ mixing we conclude that they do not significantly alter
the Standard Model result, which is dominated by the tree-level coefficient
$c_2(m_w)$.

\section{Acknowledgments}
This work was partially supported by CNPq and FAPESP.

%
\begin{table}
\caption{Values of Wilson coefficients $c_7^{TC}(m_w)$ and
 $c_8^{TC}(m_w)$ arising from colored pseudo-Nambu-Goldstone
bosons of a 1-family technicolor model for different values of
the boson mass. }
\end{table}

\begin{table}
\caption{Upper values of Wilson coefficients $c_7^{I}(m_w)$ and
 $c_8^{I}(m_w)$ in a two Higgs doublet model of type I
for different values of $\tan \beta$ and $m_{H^\pm}$. }
\end{table}

\begin{table}
\caption{Upper values of Wilson coefficients $c_7^{II}(m_w)$ and
 $c_8^{II}(m_w)$ in a two Higgs doublet model of type II
for different values of $\tan \beta$ and $m_{H^\pm}$. }
\end{table}

\newpage

\begin{center}
\begin{tabular}{|c|l|l|}  \hline
$m_{P_8^{\pm}}$ (GeV)    &  $c_7^{TC}(m_w)$    & $c_8^{TC}(m_w)$ \\  \hline
\hline
$80$          &  $-1.6 \times 10^{-2}$  &  $ -1.3 \times 10^{-2}$ \\   \hline
$400$          &  $-2.7 \times 10^{-3}  $  &  $- 4.5 \times 10^{-3}$ \\  \hline
\end{tabular}

Table I
\end{center}

\vspace{2cm}

\begin{center}
\begin{tabular}{|c|c|l|l|}  \hline
$\tan \beta$  &$m_{H^{\pm}}$ (GeV)  &$c_7^I(m_w)$   &$c_8^I(m_w)$ \\
\hline  \hline
$0.2$  &$800$  &  $-2.7 \times 10^{-3}$  &  $ -3.7 \times 10^{-3}$ \\   \hline
$0.3$  &$400$  &  $-2.8 \times 10^{-3}$  &  $- 3.6 \times 10^{-3}$ \\  \hline
\end{tabular}

Table II
\end{center}

\vspace{2cm}

\begin{center}
\begin{tabular}{|c|c|l|l|}  \hline
$\tan \beta$  &$m_{H^{\pm}}$ (GeV)  &$c_7^{II}(m_w)$   &$c_8^{II}(m_w)$ \\
\hline  \hline
$0.2$  &$800$  &  $1.2 \times 10^{-3}$  &  $ 8.7 \times 10^{-4}$ \\   \hline
$0.5$  &$300$  &  $1.5 \times 10^{-3}$  &  $ 1.1 \times 10^{-3}$ \\  \hline
$1.0$  &$200$  &  $1.7 \times 10^{-3}$  &  $ 1.3 \times 10^{-3}$ \\  \hline
\end{tabular}

Table III
\end{center}

\end{document}